\documentstyle[aps,prd,epsf]{revtex}

\newcommand{\pslash}{\not{\! p}}
\newcommand{\Pslash}{\not{\! P}}

\newcommand{\bfGamma}{{\bf \Gamma}}

\newcommand{\bfpi}{\bbox{\pi}}
\newcommand{\bfsig}{\bbox{\sigma}}
\newcommand{\bfeta}{\bbox{\eta}}

\newcommand{\be}{\begin{equation}}
\newcommand{\ee}{\end{equation}}
\newcommand{\ba}{\begin{eqnarray}}
\newcommand{\ea}{\end{eqnarray}}

\preprint{UCTP-108-00}

\begin{document}
\draft

\title{Bethe-Salpeter equation for diquarks in
color-flavor locked phase of cold dense QCD}

\author{V. A. Miransky}
\address{Bogolyubov Institute for Theoretical Physics,
         252143, Kiev, Ukraine\\
     and Research Center for Nuclear Physics,
         Osaka University,
         Osaka 567-0047,
         Japan}

\author{I. A. Shovkovy\thanks{On leave of absence from
                     Bogolyubov Institute for Theoretical
                     Physics, 252143, Kiev, Ukraine.}}
\address{Physics Department, University of Cincinnati,
         Cincinnati, Ohio 45221-0011\\
     and School of Physics and Astronomy, University of Minnesota,
         Minneapolis, MN 55455}

\author{L. C. R. Wijewardhana}
\address{Physics Department, University of Cincinnati,
         Cincinnati, Ohio 45221-0011\\
         and
         Institute of Fundamental Studies, Kandy, Sri Lanka}

\date{\today}
\maketitle

\begin{abstract} 
The Bethe-Salpeter equations for diquarks with the quantum numbers of the
Nambu-Goldstone bosons are analyzed in the color-flavor locked phase of
cold dense QCD with three quark flavors. The decay constants and the
velocities of the Nambu-Goldstone bosons are calculated in the
Pagels-Stokar approximation.  It is also shown that, in contrast to the
case of dense QCD with two flavors, there are no massive radial
excitations with quantum numbers of the Nambu-Goldstone bosons in the
color-flavor locked phase. The role of the Meissner effect in the pairing
dynamics of diquarks is explained. 
\end{abstract}

\pacs{11.10.St, 11.15.Ex, 12.38.Aw, 21.65.+f}



\section{Introduction}

Quark matter at high density is a color superconductor. It has been the
subject of many studies for the last few years. This recent increased
activity was initiated by the observation \cite{W1,S1} that the color
superconducting order parameter could be much larger than previously
thought (for old studies, see Refs.~\cite{BarFra,Bail}). Since then,
many new results have appeared
\cite{PR1,Son,us,SW2,PR2,H1,Br1,CFL,cnt,us3,us2,Spt,013,Br2,Rsch,CaD}.

Because of asymptotic freedom, QCD becomes a weakly interacting theory at
high densities. This allows one to obtain some rigorous results for dense
quark matter in the asymptotic limit. Of course, from the viewpoint of
phenomenology, it is most desirable to have a theory valid at intermediate
densities that could be produced in heavy ion collisions or could exist in
nature (for example, inside neutron or quark stars). This dilemma is
partially resolved by studying predictions of the theory at high densities
and, then, extending their validity as far as possible to the region of
interest \cite{PR1,Son,us,SW2,PR2,H1,Br1}. Notice that all the heavy quark
flavors could be safely omitted from the analysis when probing the quark
matter at realistic intermediate densities. As a result, one arrives at a
model of dense QCD with either two (``up" and ``down") or three (``up",
``down" and ``strange") flavors.

In this paper, we deal with the problem of the diquarks with the quantum
numbers of the Nambu-Goldstone (NG) bosons in cold dense QCD with three
flavors. (A similar problem in the case of two flavors was considered in
Refs.~\cite{bs-short,bs-long}.) The ground state of dense quark matter
with three flavors is a color superconductor in the so called color-flavor
locked (CFL) phase \cite{CFL}. It is remarkable that the chiral symmetry
in such a phase is broken and most of quantum numbers of physical states
coincide with those in the hadronic phase. It was tempting, therefore, to
suggest that there might exist a continuity between the two phases
\cite{cnt}.

The dynamical symmetry breaking is caused by the celebrated Cooper
instability in the pairing dynamics of quasiparticles around the Fermi
surface. As a result, the original gauge symmetry $SU(3)_{c}$ and the
global chiral symmetry $SU(3)_{L} \times SU(3)_{R}$ break down to the
global diagonal $SU(3)_{c+L+R}$ subgroup \cite{CFL}. Out of total sixteen
(would be) NG bosons, eight are removed from the physical spectrum by the
Higgs mechanism, providing masses to eight gluons. The other eight NG
bosons show up as an octet [under the unbroken $SU(3)_{c+L+R}$] of
physical particles. In addition, the global baryon number symmetry as well
as the approximate $U(1)_{A}$ symmetry also get broken. As a result, an
extra NG boson and a pseudo-NG boson appear in the low energy spectrum.
These particles are both singlets under $SU(3)_{c+L+R}$.

The low energy effective theory in the CFL phase was formulated in
Refs.~\cite{CasGat,SonSt,Rho,HZB,Zar,Beane}. Moreover, all parameters that
define the low energy action of the (pseudo-) NG bosons were calculated in
the limit of the asymptotically large chemical potential. The elegant
method of Refs.~\cite{SonSt,Rho,HZB,Zar,Beane} is based on matching some
vacuum properties (such as vacuum energy and gluon screening) of the
effective and microscopic theories. While being rather efficient for some
purposes, such a method is very limited when it comes to determining the
spectrum of bound states other than NG bosons. The other approach used to
study (diquark) bound states is based on the Bethe-Salpeter (BS) equation.
In particular, the analysis of the BS equation seems to be the only way to
test the conjecture of Ref.~\cite{us3} about the existence of an infinite
tower of massive diquark states in the color superconducting phase of
dense quark matter. In two flavor QCD, such a conjecture proved to be
partially true \cite{bs-short,bs-long}. Here we generalize the method of
Refs.~\cite{bs-short,bs-long} to the case of the CFL phase in QCD with
three flavors.

This paper is organized as follows. In Sec.~\ref{model}, we describe the
model and introduce the notation. Then, in Sec.~\ref{SD-equation}, we
review the approach of the Schwinger-Dyson equation in the color
superconducting phase of three flavor QCD. In Sec.~\ref{Ward-id}, we
derive the Ward identities for the quark-gluon vertex functions,
corresponding to the broken generators of global and gauge symmetries. We
outline the general derivation of the Bethe-Salpeter equation and present
its detailed analysis for the diquark NG bosons in Sec.~\ref{Deriv-BS-eq}.
The decay constants and velocities of the NG bosons are calculated in
Sec.~\ref{decay-consts}. In Sec.~\ref{mass}, we discuss the problem of
massive radial excitations in the diquark channel with quantum numbers of
the NG bosons. Finally, our conclusions are given in
Sec.~\ref{conclusion}.

\section{Model and notation}
\label{model}

As we mentioned in the Introduction, the original $SU(3)_{c} \times
SU(3)_{L} \times SU(3)_{R}$ symmetry of three flavor dense QCD breaks
down to the global diagonal $SU(3)_{c+L+R}$ subgroup. The condensate
in the CFL phase is given by the vacuum expectation value of the
following diquark (antidiquark) field \cite{CFL}:
\be
\langle 0|  \left(\bar{\Psi}_{D}\right)^{a}_{i}
\gamma^{5} \left(\Psi_{D}^{C}\right)^{b}_{j} |0\rangle
=\kappa_{1} \delta^{a}_{i} \delta^{b}_{j}
+\kappa_{2} \delta^{a}_{j} \delta^{b}_{i}  ,
\label{order-par}
\ee
where $\Psi_{D}$ and $\Psi_{D}^{C}=C\bar{\Psi}^{T}_{D}$ are the Dirac
spinor and its charge conjugate spinor, and $C$ is a unitary matrix
that satisfies $C^{-1} \gamma_{\mu} C=-\gamma^{T}_{\mu} $ and
$C=-C^{T}$. The complex scalar functions $\kappa_{1}$ and $\kappa_{2}$
are determined by dynamics. Here we explicitly display the flavor
($i,j=1,2,3$) and color ($a,b=1,2,3$) indices of the spinor fields.
Notice, however, that the CFL phase mixes color and flavor
representations. As a result, the notions of ``color" and ``flavor"
become essentially indistinguishable in the vacuum. In passing, we
also note that the order parameter in Eq.~(\ref{order-par}) is even
under parity.

By following Ref.~\cite{SonSt}, we introduce the color-flavor locked
Weyl spinors [octets and singlets under $SU(3)_{c+L}$ and
$SU(3)_{c+R}$, respectively] to replace the spinors of a definite
color and flavor,
\begin{eqnarray}
\psi^{A}=\frac{1}{\sqrt{2}} {\cal P}_{+} (\Psi_{D})^{~i}_{a}
\left(\lambda^{A}\right)_{i}^{~a}, &\qquad&
\psi=\frac{1}{\sqrt{3}} {\cal P}_{+} (\Psi_{D})^{~i}_{a}
\delta_{i}^{~a}, \label{def-psi} \\
\tilde{\psi}^{A}=\frac{1}{\sqrt{2}} {\cal P}_{-}
(\Psi_{D}^{C})_{j}^{~b} \left(\lambda^{A}\right)_{b}^{~j}, &\qquad&
\tilde{\psi}=\frac{1}{\sqrt{3}} {\cal P}_{-}
(\Psi_{D}^{C})_{j}^{~b} \delta_{b}^{~j}, \label{def-psi-C} \\
\phi^{A}=\frac{1}{\sqrt{2}} {\cal P}_{-} (\Psi_{D})^{~i}_{a}
\left(\lambda^{A}\right)_{i}^{~a}, &\qquad&
\phi=\frac{1}{\sqrt{3}} {\cal P}_{-} (\Psi_{D})^{~i}_{a}
\delta_{i}^{~a}, \label{def-phi} \\
\tilde{\phi}^{A}=\frac{1}{\sqrt{2}} {\cal P}_{+}
(\Psi_{D}^{C})_{j}^{~b} \left(\lambda^{A}\right)_{b}^{~j}, &\qquad&
\tilde{\phi}=\frac{1}{\sqrt{3}} {\cal P}_{+}
(\Psi_{D}^{C})_{j}^{~b} \delta_{b}^{~j}, \label{def-phi-C}
\end{eqnarray}
where $A=1,\ldots,8$, the sum over repeated indices is understood, and
${\cal P}_{\pm}=(1 \pm \gamma^5)/2$ are the left- and right-handed
projectors. Tilde denotes the charge conjugate spinors.

The order parameter in Eq.~(\ref{order-par}) is recovered by
assuming that the following (singlet under the locked symmetry)
vacuum expectation values are non-zero:
\ba
\langle 0| \bar{\psi} \tilde{\psi} |0\rangle =
-\langle 0| \bar{\phi} \tilde{\phi} |0\rangle =
-\frac{1}{2} \left(3\kappa_{1} +\kappa_{2} \right) ,
\label{ord-par1} \\
\langle 0| \bar{\psi}^{A} \tilde{\psi}^{B} |0\rangle =
-\langle 0| \bar{\phi}^{A} \tilde{\phi}^{B} |0\rangle =
-\frac{1}{2} \delta^{AB} \kappa_{2} .
\label{ord-par2}
\ea
A specific vacuum alignment leads to a relation between $\kappa_{1}$
and $\kappa_{2}$. For example, if the condensate were pure
antitriplet in color and antitriplet in flavor (with respect to the
original symmetries of the action), it would imply that
$\kappa_{2}=-\kappa_{1}$. Similarly, in the case of a sextet in color
and sextet in flavor condensate, the relation would read
$\kappa_{2}=\kappa_{1}$. It is known, however, that the true vacuum
alignment is such that the antitriplet-antitriplet contribution
dominates. At the same time, the sextet-sextet contribution is small
but non-zero \cite{CFL,us2,Spt}.

\section{Schwinger-Dyson equation}
\label{SD-equation}

In this section we briefly review the method of Schwinger-Dyson (SD) 
equation using our notation. This would also serve us as a convenient
reference point when we discuss the Bethe-Salpeter (BS) equations
in Sec.~\ref{Deriv-BS-eq}.

To start with, let us introduce the multi-component
spinor,
\be
\left(\begin{array}{c}
\Psi \\
\Psi^{A}
\end{array}\right),
\label{multi-com}
\ee
built of the left-handed Majorana spinors,
\ba
\Psi &=& \psi+\tilde{\psi}  \label{def-Maj},\\
\Psi^{A} &=& \psi^{A}+\tilde{\psi}^{A},
\label{def-MajA}
\ea
where $A=1,\ldots,8$. Similarly, we could introduce the multi-spinors made
of the right-handed Majorana fields, $\Phi$ and $\Phi^{A}$. In our
analysis, restricted only to the (hard dense loop improved) rainbow
approximation, the left and right sectors of the theory completely
decouple. Then, without loss of generality, it is sufficient to study the
SD equation only in one of the sectors.

The benefit of using the notation in Eq.~(\ref{multi-com}) is that
the inverse full propagator of quarks takes a very simple
block-diagonal form,
\ba
G^{-1}(p)&=&\left(\begin{array}{cc}
S^{-1}_{1}(p) & 0 \\
0 & \delta^{AB} S^{-1}_{2}(p) \end{array}\right), \label{G-1} \\
S^{-1}_{1}(p)&=&-i\left(\not{\! p} +\mu \gamma^0 \gamma^5
+ \Delta_{1}{\cal P}_{-} +\tilde{\Delta}_{1} {\cal P}_{+}\right),
\label{S1-1} \\
S^{-1}_{2}(p)&=&-i\left( \not{\! p} +\mu \gamma^0 \gamma^5
-\Delta_{2} {\cal P}_{-} -\tilde{\Delta}_{2} {\cal P}_{+} \right) .
\label{S2-1}
\ea
Here  $\Delta_{1,2} = \Delta^{+}_{1,2} \Lambda^{+}_{p}  +
\Delta^{-}_{1,2}\Lambda^{-}_{p}$ and  $\tilde{\Delta}_{1,2}=\gamma^0
\Delta^{\dagger}_{1,2}\gamma^0$, and the ``on-shell"  projectors of
quarks,
\ba
\Lambda_{p}^{\pm }=\frac{1}{2}
\left(1\pm \frac{\vec{\alpha} \cdot \vec{p}}{|\vec{p}|}\right),
\quad \vec{\alpha} =\gamma^{0} \vec{\gamma} ,
\ea
are the same as in Ref.~\cite{us2}. We note that the gaps,
$\Delta_{1}$ and $\Delta_{2}$, enter the propagators (\ref{S1-1})
and (\ref{S2-1}) with opposite signs.

After expressing the standard bare vertex of QCD in terms of Majorana
spinors (\ref{def-Maj}) and (\ref{def-MajA}), we arrive at the
following matrix form of the vertex:
\be
\gamma^{A\mu}=\gamma^{\mu}\left(\begin{array}{cc}
0 & \frac{1}{\sqrt{6}}\delta^{AC} \gamma^{5} \\
\frac{1}{\sqrt{6}}\delta^{AB} \gamma^{5} &
\frac{1}{2}( d^{ABC} \gamma^{5} -i f^{ABC})
\end{array}\right),
\label{vertex}
\ee
where $d^{ABC}$ and $f^{ABC}$ are defined by the (anti-) commutation
relations of the Gell-Mann matrices,
\ba
\left\{ \lambda^{A}, \lambda^{B} \right\} &=& \frac{4}{3} \delta^{AB}
+ 2 d^{ABC} \lambda^{C} , \\
\left[ \lambda^{A}, \lambda^{B} \right] &=&
2i f^{ABC} \lambda^{C} .
\ea

With all the ingredients at hand, it is straightforward to derive the
matrix form of the SD equation,
\be
G^{-1}(p) = G^{-1}_{0}(p)
+4\pi\alpha_{s}\int\frac{d^4 q}{(2\pi)^4}
\gamma^{A\mu} G(q) \Gamma^{A\nu}(q,p)
{\cal D}_{\mu\nu}(q-p).
\label{SD-eq}
\ee
Here $\gamma^{A\mu}$ and $\Gamma^{A\mu}$ are the bare and the full
vertices, respectively. The gluon propagator in the SD equation is
the same as in Ref.~\cite{us2}, except that the Meissner effect
should, in general, be taken into account.

By inverting the expression in Eq.~(\ref{G-1}), we obtain the
following representation for the quark propagator:
\be
G(p)=\left(\begin{array}{cc}
S_{1}(p) & 0 \\
0 & \delta^{AB} S_{2}(p)
\end{array}\right),
\label{propagator-L}
\ee
where
\ba
S_{1}(p) &=&
i\frac{\gamma^0 (p_0+\epsilon_{p}^{+})-\Delta^{+}_{1} }
{p_0^2-(\epsilon_{p}^{+})^2-|\Delta^{+}_{1} |^2}
\Lambda_{p}^{-} {\cal P}_{+}
+i\frac{\gamma^0 (p_0-\epsilon_{p}^{+})-(\Delta^{+}_{1} )^{*}}
{p_0^2-(\epsilon_{p}^{+})^2-|\Delta^{+}_{1} |^2}
\Lambda_{p}^{+} {\cal P}_{-} \nonumber \\
&+& i\frac{\gamma^0 (p_0-\epsilon_{p}^{-})-\Delta^{-}_{1} }
{p_0^2-(\epsilon_{p}^{-})^2-|\Delta^{-}_{1} |^2}
\Lambda_{p}^{+} {\cal P}_{+}
+i\frac{\gamma^0 (p_0+\epsilon_{p}^{-})-(\Delta^{-}_{1} )^{*}}
{p_0^2-(\epsilon_{p}^{-})^2-|\Delta^{-}_{1} |^2}
\Lambda_{p}^{-} {\cal P}_{-} , \label{S-1} \\
S_{2}(p) &=&
i\frac{\gamma^0 (p_0+\epsilon_{p}^{+})+\Delta^{+}_{2} }
{p_0^2-(\epsilon_{p}^{+})^2-|\Delta^{+}_{2} |^2}
\Lambda_{p}^{-} {\cal P}_{+}
+i\frac{\gamma^0 (p_0-\epsilon_{p}^{+})+(\Delta^{+}_{2} )^{*}}
{p_0^2-(\epsilon_{p}^{+})^2-|\Delta^{+}_{2} |^2}
\Lambda_{p}^{+} {\cal P}_{-} \nonumber \\
&+& i\frac{\gamma^0 (p_0-\epsilon_{p}^{-})+\Delta^{-}_{2} }
{p_0^2-(\epsilon_{p}^{-})^2-|\Delta^{-}_{2} |^2}
\Lambda_{p}^{+} {\cal P}_{+}
+i\frac{\gamma^0 (p_0+\epsilon_{p}^{-})+(\Delta^{-}_{2} )^{*}}
{p_0^2-(\epsilon_{p}^{-})^2-|\Delta^{-}_{2} |^2}
\Lambda_{p}^{-} {\cal P}_{-} . \label{S-2}
\ea
The bare propagator in Eq.~(\ref{SD-eq}) is similar but with zero
values of the gaps $\Delta^{\pm}_{n}$ ($n=1,2$).

In the (hard dense loop improved) rainbow approximation, both vertices
in the SD equation are bare. By making use of the propagator in
Eq.~(\ref{propagator-L}) and the vertex in Eq.~(\ref{vertex}),
we derive the set of two gap equations \cite{us2,Spt},
\ba
\Delta_{1}(p) &=& -\frac{16}{3}\pi\alpha_{s}\int\frac{d^4 q}{(2\pi)^4}
\gamma^{\mu} {\cal P}_{+} S_{2}(q) {\cal P}_{+} \gamma^{\nu}
 {\cal D}_{\mu\nu}(q-p), \label{D-1} \\
\Delta_{2}(p) &=& \frac{2}{3}\pi\alpha_{s}\int\frac{d^4 q}{(2\pi)^4}
\gamma^{\mu} {\cal P}_{+} \left[ 2 S_{2}(q) - S_{1}(q) \right]
{\cal P}_{+} \gamma^{\nu}{\cal D}_{\mu\nu}(q-p),
\label{D-2}
\ea
where we used the identities
\ba
d^{ACD} d^{BCD} = \frac{5}{3} \delta^{AB}, \\
f^{ACD} f^{BCD} = 3 \delta^{AB}.
\ea
An approximate analytical solution to this coupled system of gap
equations was presented in Ref.~\cite{us2} (a numerical solution was
also given in Ref.~\cite{Spt}). Here we quote the final results
\cite{us2},
\begin{equation}
\Delta^{-}_{1}(p_{4})\approx 2\Delta^{-}_{2}(p_{4})
\approx 2\Delta^{-}_{(\bar{3},\bar{3})}(p_{4}),
\label{D1-D2}
\end{equation}
where
\begin{mathletters}
\begin{eqnarray}
\Delta^{-}_{(\bar{3},\bar{3})}(p_{4}) &\approx &\Delta_{0},
\quad p_{4}\leq \Delta_{0},\\
\Delta^{-}_{(\bar{3},\bar{3})}(p_{4})&\approx&\Delta_{0}
\sin\left(\sqrt{ \frac{2\alpha_{s}}{9\pi} }
\ln\frac{\Lambda}{p_{4}}\right),
\quad  p_{4}\geq \Delta_{0},
\end{eqnarray}
\label{sigma_1}
\end{mathletters}
with $\Lambda \equiv 16(2\pi)^{3/2}\mu/(3\alpha_{s})^{5/2}$ and
\begin{equation}
\Delta_{0}\approx
\frac{16(2\pi)^{3/2}\mu}{(3\alpha_{s})^{5/2}}
\exp\left(-\frac{3\pi^{3/2}}{2^{3/2}\sqrt{\alpha_{s}}}\right).
\end{equation}
As in the case of two flavor dense QCD, the issue of the overall
constant factor in this expression is still unsettled. Some sources of
possible corrections are discussed in Refs.~\cite{Br1,013}. In
addition, we could argue (along the same lines as in Appendix~B of
Ref.~\cite{bs-long}) that there is also at least one non-perturbative
correction that could modify the constant factor in the expression for
the gap.

\section{Ward identities}
\label{Ward-id}

In order to preserve the gauge invariance in dense QCD, one has to
make sure that Green functions satisfy the Ward identities. In this
section, we consider the simplest Ward identities that relate the
vertex functions and the quark propagators. In addition to
establishing the longitudinal part of the full vertex function, these
identities will play a very important role in our analysis of the BS
equations for the NG bosons.

In general, the structure of Ward identities in non-Abelian gauge
theories (the Slavnov-Taylor identities) is much more complicated than
in Abelian ones: they include contributions of the  Faddeev-Popov
ghosts. Fortunately in the (hard dense loop improved) ladder
approximation, used in this paper, the situation simplifies. Indeed,
since the direct interactions between gluons are neglected in this
approximation, the Ward identities have an Abelian-like structure.

To start with, let us rewrite the conserved currents (related to the
baryon number and color symmetry) in terms of the Weyl spinors, defined in
Eqs.~(\ref{def-Maj}) and (\ref{def-MajA}). In this approximation, the
left- and right-handed sectors of the theory decouple and the axial
$U(1)_A$ charge is conserved. Therefore we can consider the
(approximately) conserved currents in the two sectors separately. Here we
give the details of the analysis for the left sector. The expressions for
the singlet (related to the baryon number and the axial charge) and the
octet (related to the color charge) channels read
\ba
j_{\mu}(x) &=& \bar{\Psi}_{D}(x) \gamma_{\mu} {\cal P}_{+}
\Psi_{D}(x) = \frac{1}{2}
\bar{\Psi}^{A}(x) \gamma_{\mu} \gamma^{5} \Psi^{A}(x)
+\frac{1}{2}
\bar{\Psi} (x) \gamma_{\mu} \gamma^{5} \Psi (x) ,
\label{current-L} \\
j^{A}_{\mu}(x) &=& \frac{1}{2} \bar{\Psi}_{D}(x) \gamma_{\mu}
\lambda^{A} {\cal P}_{+} \Psi_{D}(x) \nonumber\\
&=& \frac{1}{4} \bar{\Psi}^{B}(x) \gamma_{\mu}
\left(d^{ABC} \gamma^{5} -if^{ABC}\right)
\Psi^{C}(x) +\frac{1}{2\sqrt{6}} \left[\bar{\Psi}(x) \gamma_{\mu}
\gamma^{5} \Psi^{A}(x) +\bar{\Psi}^{A}(x) \gamma_{\mu} \gamma^{5}
\Psi(x) \right] .
\label{current}
\ea
Similar expressions could be written in the right-handed sector too.
Now, we are interested in the Ward identities that relate the
quark-gluon vertices to the propagators of quarks. Therefore, let us
introduce the following (non-amputated) vertex functions:
\begin{mathletters}
\ba
\bfGamma_{\mu}(x,y) &=& \langle 0 | T
j_{\mu}(0) \Psi(x) \bar{\Psi}(y) | 0 \rangle
, \label{ver-0} \\
\bfGamma^{AB}_{\mu}(x,y) &=& \langle 0 | T
j_{\mu}(0) \Psi^{A}(x) \bar{\Psi}^{B}(y) | 0 \rangle
, \label{ver-00} \\
\bfGamma^{A,BC}_{\mu}(x,y) &=& \langle 0 | T
j^{A}_{\mu}(0) \Psi^{B}(x) \bar{\Psi}^{C}(y) | 0 \rangle
, \label{ver-1} \\
\bfGamma^{A,B}_{1,\mu}(x,y) &=& \langle 0 | T
j^{A}_{\mu}(0) \Psi^{B}(x) \bar{\Psi}(y) | 0 \rangle
, \label{ver-2}  \\
\bfGamma^{A,B}_{2,\mu}(x,y) &=& \langle 0 | T
j^{A}_{\mu}(0) \Psi(x) \bar{\Psi}^{B}(y) | 0 \rangle
. \label{ver-3}
\ea
\label{vertices}
\end{mathletters}
In order to derive the Ward identities, one needs to know the
transformation properties of the quark fields. By making use of the
transformation properties of the Dirac spinors, it is straightforward
to derive the following infinitesimal baryon conservation symmetry
transformations for the spinors of interest:
\ba
\delta \Psi^{A} &=& i \omega \gamma^{5} \Psi^{A},\label{del-p-A}\\
\delta \bar{\Psi}^{A} &=& i \omega \bar{\Psi}^{A} \gamma^{5},
\label{del-b-A}\\
\delta \Psi &=& i \omega \gamma^{5} \Psi, \label{del-p} \\
\delta \bar{\Psi} &=& i \omega \bar{\Psi} \gamma^{5} ,
\label{del-b}
\ea
as well as the following color symmetry transformations:
\ba
\delta \Psi^{A} &=& \frac{i}{2}
\omega^{B} \left( d^{ABC} \gamma^{5} +i f^{ABC} \right) \Psi^{C}
+\frac{i\omega^{A}}{\sqrt{6}} \gamma^{5} \Psi ,\label{del-psi-A}\\
\delta \bar{\Psi}^{A} &=& \frac{i}{2} \omega^{B} \bar{\Psi}^{C}
\left( d^{ABC} \gamma^{5} +i f^{ABC} \right)
+\frac{i\omega^{A}}{\sqrt{6}} \bar{\Psi} \gamma^{5},
\label{del-bar-A}\\
\delta \Psi &=& \frac{i\omega^{A}}{\sqrt{6}}
\gamma^{5} \Psi^{A}, \label{del-psi} \\
\delta \bar{\Psi} &=& \frac{i\omega^{A}}{\sqrt{6}}
\bar{\Psi}^{A} \gamma^{5} . \label{del-bar}
\ea
Now, by making use of the current conservation as well as the
definition of the vertices in Eq.~(\ref{vertices}), we
straightforwardly derive the Ward identities for the non-amputated
vertices:
\begin{mathletters}
\ba
P^{\mu} \bfGamma_{\mu}\left(k+P,k\right)
&=& i \left[ \gamma^{5} S_{1}\left(k\right) + S_{1}\left(k+P\right)
\gamma^{5} \right]
,\label{27} \\
P^{\mu} \bfGamma^{AB}_{\mu}\left(k+P,k\right)
&=& i \delta^{AB}
\left[ \gamma^{5} S_{2}\left(k\right) + S_{2}\left(k+P\right) \gamma^{5}
\right]
,\label{27d} \\
P^{\mu} \bfGamma^{A,BC}_{\mu}\left(k+P,k\right)
&=& \frac{i}{2} d^{ABC}
\left[ \gamma^{5} S_{2}\left(k\right) + S_{2}\left(k+P\right) \gamma^{5}
\right] \nonumber\\
&&-\frac{1}{2} f^{ABC} \left[S_{2}\left(k\right) - S_{2}\left(k+P\right)
\right]
,\label{27a} \\
P^{\mu} \bfGamma^{A,B}_{1,\mu}\left(k+P,k\right)
&=& \frac{i}{\sqrt{6}} \delta^{AB}
\left[ \gamma^{5} S_{1}\left(k\right) + S_{2}\left(k+P\right) \gamma^{5}
\right]
,\label{27b}\\
P^{\mu} \bfGamma^{A,B}_{2,\mu}\left(k+P,k\right)
&=& \frac{i}{\sqrt{6}} \delta^{AB}
\left[ \gamma^{5} S_{2}\left(k\right) + S_{1}\left(k+P\right) \gamma^{5}
\right]
,\label{27c}
\ea
\label{Ward-LR}
\end{mathletters}
where $S_{1}$ and $S_{2}$ are the quark propagators. In the leading
order approximation where the wave function renormalization
corrections are neglected, the explicit form of the momentum space
propagators is given in Eqs.~(\ref{S-1}) and (\ref{S-2}).

At this point, let us note that the use of the non-amputated vertices
was crucial for the derivation of the Ward identities. Other than
that, the non-amputated vertices are not very convenient to work with.
In fact, it is the amputated rather than the non-amputated vertices
that are usually used in the Feynman diagrams. Similarly, it is the
amputated vertices that will appear in the BS equation in
Sec.~\ref{Deriv-BS-eq}. The formal definitions of the amputated
vertices read
\begin{mathletters}
\ba
\Gamma_{\mu}\left(k+P,k\right) &=&
  S^{-1}_{1}\left(k+P\right) \bfGamma_{\mu}
\left(k+P,k\right) S^{-1}_{1}
\left(k\right)  ,
\label{30}\\
\Gamma^{AB}_{\mu}\left(k+P,k\right) &=&
  S^{-1}_{2}\left(k+P\right) \bfGamma^{AB}_{\mu}\left(k+P,k\right)
S^{-1}_{2}\left(k\right) ,
\label{30d}\\
\Gamma^{A,BC}_{\mu}\left(k+P,k\right) &=&
  S^{-1}_{2}\left(k+P\right) \bfGamma^{A,BC}_{\mu}\left(k+P,k\right)
S^{-1}_{2}\left(k\right) ,
\label{30a}\\
\Gamma^{A,B}_{1,\mu}\left(k+P,k\right) &=&
  S^{-1}_{2}\left(k+P\right) \bfGamma^{A,B}_{1,\mu}\left(k+P,k\right)
S^{-1}_{1}\left(k\right) ,
\label{30b}\\
\Gamma^{A,B}_{2,\mu}\left(k+P,k\right) &=&
  S^{-1}_{1}\left(k+P\right) \bfGamma^{A,B}_{2,\mu}\left(k+P,k\right)
S^{-1}_{2}\left(k\right) .
\label{30c}
\ea
\label{amp-def}
\end{mathletters}
It follows from
Eq.~(\ref{Ward-LR}) that they satisfy the following identities of
their own:
\begin{mathletters}
\ba
P^{\mu} \Gamma_{\mu}\left(k+P,k\right)
&=& i \left[ S^{-1}_{1}\left(k+P\right) \gamma^{5}
+ \gamma^{5} S^{-1}_{1}\left(k\right) \right],
\label{a} \\
P^{\mu} \Gamma^{AB}_{\mu}\left(k+P,k\right)
&=& i \delta^{AB} \left[ S^{-1}_{2}\left(k+P\right) \gamma^{5}
+ \gamma^{5} S^{-1}_{2}\left(k\right) \right],
\label{d} \\
P^{\mu} \Gamma^{A,BC}_{\mu}\left(k+P,k\right)
&=& \frac{i}{2} d^{ABC}
\left[ S^{-1}_{2}\left(k+P\right) \gamma^{5}
+ \gamma^{5} S^{-1}_{2}\left(k\right) \right] \nonumber\\
&&-\frac{1}{2} f^{ABC} \left[S^{-1}_{2}\left(k+P\right)
- S^{-1}_{2}\left(k\right) \right],
\label{aa} \\
P^{\mu} \Gamma^{A,B}_{1,\mu}\left(k+P,k\right)
&=& \frac{i}{\sqrt{6}} \delta^{AB}
\left[ S^{-1}_{2}\left(k+P\right) \gamma^{5}
+ \gamma^{5} S^{-1}_{1}\left(k\right) \right], \label{ab}\\
P^{\mu} \Gamma^{A,B}_{2,\mu}\left(k+P,k\right)
&=& \frac{i}{\sqrt{6}} \delta^{AB}
\left[ S^{-1}_{1}\left(k+P\right) \gamma^{5}
+ \gamma^{5} S^{-1}_{2}\left(k\right) \right] .\label{ac}
\ea
\label{Ward-amp}
\end{mathletters}
In the rest of the paper, we are going to use these Ward identities a
number of times. Because of a relatively simple structure of the
inverse quark propagators, the last form of the identities for the
amputated vertices will be particularly convenient.

Notice that in the limit $P\to 0$, we obtain
\begin{mathletters}
\ba
\left. P^{\mu} \Gamma_{\mu}\left(k+P,k\right) \right|_{P\to 0}
&=& 2 \left( \tilde{\Delta}_{1}(k) {\cal P}_{+}
-\Delta_{1}(k) {\cal P}_{-} \right), \label{P1=0} \\
\left. P^{\mu} \Gamma^{AB}_{\mu}\left(k+P,k\right) \right|_{P\to 0}
&=& 2 \delta^{AB} \left( \Delta_{2}(k) {\cal P}_{-}
-\tilde{\Delta}_{2}(k) {\cal P}_{+} \right), \label{P2=0} \\
\left.  P^{\mu} \Gamma^{A,BC}_{\mu}\left(k+P,k\right) \right|_{P\to 0}
&=& d^{ABC} \left( \Delta_{2}(k) {\cal P}_{-}
-\tilde{\Delta}_{2}(k) {\cal P}_{+}\right) , \label{P3=0} \\
\left.  P^{\mu} \Gamma^{A,B}_{1,\mu}\left(k+P,k\right) \right|_{P\to 0}
&=&
\frac{1}{\sqrt{6}}\delta^{AB} \left[ \left( \tilde{\Delta}_{1}(k)
-\tilde{\Delta}_{2}(k)\right) {\cal P}_{+}
+ \left(\Delta_{2}(k)-\Delta_{1}(k) \right) {\cal P}_{-}
\right] , \label{P4=0} \\
\left.  P^{\mu} \Gamma^{A,B}_{2,\mu}\left(k+P,k\right) \right|_{P\to 0}
&=&
\frac{1}{\sqrt{6}}\delta^{AB} \left[ \left( \tilde{\Delta}_{1}(k)
-\tilde{\Delta}_{2}(k)\right) {\cal P}_{+}
+ \left(\Delta_{2}(k)-\Delta_{1}(k) \right) {\cal P}_{-}
\right]. \label{P5=0}
\ea
\label{Ward-a}
\end{mathletters}
These expressions show that all five vertices have poles at $P=0$.
Such poles indicate the presence of NG bosons that correspond to
broken continuous symmetries in the theory.

Let us clarify the exact origin of the poles in all the vertices
introduced. It is straightforward to show that the pole contributions in
the first two vertices, defined in Eqs.~(\ref{ver-0}) and (\ref{ver-00}),
indicate the appearance of the NG boson related to breaking of the baryon
number as well as the axial symmetry: in this approximation, the
$U(1)_{A}$ charge is conserved and the axial symmetry is spontaneously
broken. This boson is a singlet with respect to the unbroken
$SU(3)_{L+R+c}$ subgroup. Similarly, one can see that the poles of the
vertices in Eqs.~(\ref{ver-1}), (\ref{ver-2}) and (\ref{ver-3}) are due to
an octet of the NG bosons related to breaking of the color symmetry
(because of the locking, the chiral symmetry is also broken). Notice that
considering the right-handed sector will double the number of the NG
bosons.

One could also establish that the explicit pole structure of the vertices
should read
\begin{mathletters}
\ba
\left. \bfGamma_{\mu}\left(k+P,k\right) \right|_{P\to 0}
&=& \frac{P^{(\eta)}_{\mu} F^{(\eta)}}{P^{(\eta)}_{\nu}P^{\nu}}
\bfeta (k,0), \label{pole0} \\
\left.\bfGamma^{AB}_{\mu}\left(k+P,k\right) \right|_{P\to 0}
&=& \frac{P^{(\eta)}_{\mu} F^{(\eta)}}{P^{(\eta)}_{\nu}P^{\nu}}
\delta^{AB} \bfeta^{\prime} (k,0), \label{pole00} \\
\left. \bfGamma^{A,BC}_{\mu}\left(k+P,k\right) \right|_{P\to 0}
&=&\frac{P^{(\pi)}_{\mu} F^{(\pi)}}{P^{(\pi)}_{\nu}P^{\nu}}
d^{ABC} \bfpi_{0}(k,0), \label{pole1}\\
\left. \bfGamma^{A,B}_{1,\mu}\left(k+P,k\right) \right|_{P\to 0}
&=& \frac{P^{(\pi)}_{\mu} F^{(\pi)}}{P^{(\pi)}_{\nu}P^{\nu}}
\delta^{AB} \bfpi_{1} (k,0), \label{pole2} \\
\left. \bfGamma^{A,B}_{2,\mu}\left(k+P,k\right) \right|_{P\to 0}
&=& \frac{P^{(\pi)}_{\mu} F^{(\pi)}}{P^{(\pi)}_{\nu}P^{\nu}}
\delta^{AB} \bfpi_{2} (k,0), \label{pole3}
\ea
\label{poles}
\end{mathletters}
where we introduced the notation $P^{(x)}_{\mu} =(P_{0}, -c_{x}^{2}
\vec{P})$ (with $c_{x}$ being the velocity of the appropriate NG
boson). The decay constants are defined by
\begin{mathletters}
\ba
\langle 0| j_{\mu}(0) | P \rangle  &=&
i P^{(\eta)}_{\mu} F^{(\eta)} , \label{decay-1} \\
\langle 0| j^{A}_{\mu}(0) | P; B \rangle &=&
i \delta^{AB} P^{(\pi)}_{\mu} F^{(\pi)} . \label{decay-8}
\ea
\label{decay}
\end{mathletters}
The BS wave functions in the coordinate representation read
\begin{mathletters}
\ba
\bfeta (x,y;P) &\equiv& e^{-iP(x+y)/2} \bfeta (x-y;P)
=\langle 0| T \Psi (x) \bar{\Psi} (y)
|P \rangle , \label{def-eta1} \\
\bfeta^{\prime} (x,y;P) &\equiv& e^{-iP(x+y)/2} \bfeta^{\prime} (x-y;P)
=\frac{1}{8} \langle 0| T \Psi^{A} (x)
\bar{\Psi}^{A} (y) |P \rangle , \label{def-eta2}
\ea
\label{def-eta}
\end{mathletters}
and
\begin{mathletters}
\ba
\bfpi_{0}^{A(B)} (x,y;P) &\equiv& \delta^{AB}
e^{-iP(x+y)/2} \bfpi_{0}(x-y;P) = \frac{3}{5} d^{ACD}
\langle 0| T \Psi^{C}(x) \bar{\Psi}^{D}(y)
|P; B \rangle  , \quad A,B=1,\ldots, 8 , \label{def-pi1} \\
\bfpi_{1}^{A(B)} (x,y;P) &\equiv& \delta^{AB}
e^{-iP(x+y)/2} \bfpi_{1}(x-y;P)
= \langle 0| T \Psi^{A}(x) \bar{\Psi}(y)
|P; B \rangle  , \quad A,B=1,\ldots, 8  , \label{def-pi2}\\
\bfpi_{2}^{A(B)} (x,y;P)  &\equiv& \delta^{AB}
e^{-iP(x+y)/2} \bfpi_{2}(x-y;P)
= \langle 0| T \Psi (x) \bar{\Psi}^{A} (y)
|P ; B \rangle  , \quad A,B=1,\ldots, 8,  \label{def-pi3}
\ea
\label{def-pi}
\end{mathletters}
in the singlet and the octet channels, respectively. The momentum
representation of the corresponding wave functions are given by the
Fourier transforms of the translation invariant parts.

\section{BS equations for NG bosons}
\label{Deriv-BS-eq}

The bound states and resonances should reveal themselves through the
appearance of poles in Green functions. To consider the problem of diquark
bound states in cold dense QCD, one has to introduce four-point Green
functions that describe the two particle scattering in the diquark
channels of interest. The residues at the poles of the Green functions are
related to the BS wave functions of the bound states. By starting from the
(inhomogeneous) BS equations for the four-point Green functions, it is
straightforward to derive the so-called homogeneous BS equations for the
wave functions.

The NG bosons in the problem at hand are the composite diquark states.
Altogether, there are seventeen NG bosons and one pseudo-NG boson [the
latter is related to breaking of the approximate $U(1)_{A}$ symmetry]. As
was already mentioned in Sec. \ref{Ward-id}, there is no difference
between the pseudo-NG boson and the real NG bosons in the ladder
approximation. By combining the NG bosons in left and right sectors of the
theory, we could construct the corresponding set of scalar and
pseudo-scalar states. We note, however, that the scalar octet is
unphysical because of the Higgs mechanism. The pseudoscalar octet is built
of real NG bosons related to breaking of chiral symmetry. In addition,
there is a scalar singlet and a pseudoscalar singlet. Both of them are
physical, but the first is a real NG boson, while the other is a pseudo-NG
boson. While considering the BS equation in the left-handed sector, we
could reveal only half of the total eighteen NG bosons. As is clear, they
should belong to a singlet and an octet representations of
$SU(3)_{L+R+c}$.

We would like to note that though the scalars from the octet are removed
from the physical spectrum by the Higgs mechanism, they exist in the
theory as "ghosts" \cite{JJ}, and one cannot get rid of them completely,
unless a unitary gauge is found. In fact these ghosts play an important
role in getting rid of unphysical poles from on-shell scattering
amplitudes \cite{JJ}. We will use covariant gauges: because of the
composite nature of the order parameter, it does not seem to be
straightforward to define and to use the unitary gauge here.

The nine Majorana quark fields in Eq.~(\ref{multi-com}) form a singlet
and an octet with respect to the color-flavor locked group of the
vacuum. Clearly, we could construct all kinds of different diquarks out
of these building blocks. If there is an attraction in the appropriate
channel, two constituent singlets could give rise to a diquark
singlet. Similarly, one singlet and one octet could produce an octet
of diquarks. Finally, out of two quark octets, one could construct
a whole new set of additional diquarks. Indeed, by making use of the
representation theory,
\be
8 \otimes 8 = 1 \oplus 8_{s} \oplus 8_{a} \oplus 10 \oplus \bar{10}
\oplus 27, \label{8times8}
\ee
we might expect six more multiplets. As was shown in the previous
section, however, only the singlet and the first of the octets (the
symmetric one) are relevant for the description of the NG bosons. One
could also guess that these singlet and octet are the most attractive
channels in the color-flavor locked phase of dense QCD. For this
reason, we restrict our study of the general bound state problem to
the analysis of only these two channels [though one should keep in
mind a possibility of a mixing between those two octets (see Sec.
\ref{BS-octet})].

In order to derive the BS equations, we use the method developed in
Ref.~\cite{bs-long}. To this end, we need to use the following
effective Lagrangian:
\ba
&& {\cal L}_{eff} = \frac{1}{2}
\bar{\Psi} \left( \pslash +\mu \gamma^0 \gamma^5
+\Delta_{1} {\cal P}_{-} + \tilde{\Delta}_{1} {\cal P}_{+} \right)
\Psi  + \frac{1}{2}
\bar{\Psi}^{A} \left( \pslash +\mu \gamma^0 \gamma^5 -
\Delta_{2}  {\cal P}_{-} - \tilde{\Delta}_{2}  {\cal P}_{+} \right)
\Psi^{A} \nonumber \\
&& +\frac{1}{4} \bar{\Psi}^{B}(x) A^{A}_{\mu} \gamma^{\mu}
\left(d^{ABC} \gamma^{5} -if^{ABC}\right)
\Psi^{C}(x) +\frac{1}{2\sqrt{6}} A^{A}_{\mu} \left(\bar{\Psi}(x)
\gamma^{\mu} \gamma^{5} \Psi^{A}(x) +\bar{\Psi}^{A}(x) \gamma^{\mu}
\gamma^{5} \Psi(x) \right),
\label{L-eff}
\ea
plus the right-handed contribution. This effective Lagrangian is a
starting point in the derivation of the BS equations for the wave
functions introduced in Eqs.~(\ref{def-eta}) and (\ref{def-pi}).
While dealing with the multicomponent spinor defined in
Eq.~(\ref{multi-com}), it is rather natural to introduce the
corresponding matrix form of the BS wave function. We use $X(p;P)$ as
a generic notation for such a matrix wave function.

In the (hard dense loop improved) ladder approximation, the matrix
form of the BS equation reads
\be
 G^{-1}\left(p+\frac{P}{2}\right)
X(p;P) G^{-1}\left(p-\frac{P}{2}\right)
=-4\pi \alpha_{s}\int\frac{d^4 q}{(2\pi)^4}
\gamma^{A\mu} X(q;P) \gamma^{B\nu}
{\cal D}^{AB}_{\mu\nu}(q-p) ,
\label{BS-matrix-eq}
\ee
where ${\cal D}^{AB}_{\mu\nu}(q-p)$ is the gluon propagator and
$\gamma^{A\mu}$ is the bare quark-gluon vertex given in
Eq.~(\ref{vertex}). This approximation has the same status as the
rainbow approximation in the SD equation. It assumes that the coupling
constant is weak, and the leading perturbative expression for the
kernel of the BS equation adequately represents the quark
interactions.

\subsection{BS equation in the singlet channel}
\label{BS-singlet}

Now, let us consider the BS equation in the singlet channel. As was
explained above, it is sufficient to consider only the left-handed sector
in this approximation. As we know from studying the pole structure of the
vertices in Sec.~\ref{Ward-id}, the singlet channel should contain at
least one bound state, the diquark NG boson related to breaking of the
baryon number [or, what is almost the same in our notation, to the
approximate $U(1)_{A}$ symmetry].

We start the derivation of the BS equation from establishing the
structure of the wave function. By incorporating the definition in
Eq.~(\ref{def-eta}), we obtain the following matrix form of the
(non-amputated) BS wave function in the singlet channel:
\be
X_{\eta}(p,P) = \left(\begin{array}{cc}
\bfeta(p,P) & 0 \\
0 & \delta^{AB} \bfeta^{\prime}(p,P)
\end{array}\right).
\label{mat-BS1wf}
\ee
Now, we substitute this into the BS equation (\ref{BS-matrix-eq}) and
arrive at the following set of equations:
\ba
S^{-1}_{1}\left(p+\frac{P}{2}\right) \bfeta (p,P)
S^{-1}_{1}\left(p-\frac{P}{2}\right) &=& \frac{16}{3}
\pi \alpha_{s} \int \frac{d^4 q}{(2\pi )^4} \gamma^{\mu}
\gamma^{5} \bfeta^{\prime} (q,P) \gamma^{5}
\gamma^{\nu} {\cal D}_{\mu\nu}(q-p), \label{bs11} \\
S^{-1}_{2}\left(p+\frac{P}{2}\right) \bfeta^{\prime} (p,P)
S^{-1}_{2}\left(p-\frac{P}{2}\right) &=&
\frac{2}{3} \pi \alpha_{s} \int \frac{d^4 q}{(2\pi )^4} \gamma^{\mu}
\left( \gamma^{5} \bfeta (q,P) \gamma^{5}
+\frac{5}{2}\gamma^{5} \bfeta^{\prime} (q,P) \gamma^{5}
-\frac{9}{2}\bfeta^{\prime}\right)
\gamma^{\nu} {\cal D}_{\mu\nu}(q-p). \label{bs22}
\ea
In order to simplify the analysis, it is rather convenient to
introduce the amputated BS wave functions,
\begin{mathletters}
\ba
\eta(p,P) &=& S^{-1}_{1}\left(p+\frac{P}{2}\right) \bfeta(p,P)
S^{-1}_{1}\left(p-\frac{P}{2}\right) ,
\label{eta-amp}\\
\eta^{\prime}(p,P) &=& S^{-1}_{2}\left(p+\frac{P}{2}\right)
\bfeta^{\prime} (p,P)
S^{-1}_{2}\left(p-\frac{P}{2}\right) .  \label{eta1-amp}
\ea
\label{eta-amp-def}
\end{mathletters}
In addition, let us restrict ourselves to the analysis of the
the BS wave functions in the limit $P \to 0$. As one could check,
such a limit is well defined and it is consistent with the on-shell
condition for the NG bosons. The BS equations, then, read
\ba
\eta (p) &=& \frac{16}{3} \pi \alpha_{s} \int \frac{d^4 q}{(2\pi )^4}
\gamma^{\mu} \gamma^{5} S_{2}(q) \eta^{\prime} (q) S_{2}(q) \gamma^{5}
\gamma^{\nu} {\cal D}_{\mu\nu}(q-p), \label{bs11-amp} \\
\eta^{\prime} (p) &=&
\frac{2}{3} \pi \alpha_{s} \int \frac{d^4 q}{(2\pi )^4} \gamma^{\mu}
\left( \gamma^{5} S_{1}(q) \eta (q) S_{1}(q) \gamma^{5}
+\frac{5}{2}\gamma^{5} S_{2}(q)
\eta^{\prime} (q) S_{2}(q) \gamma^{5}
-\frac{9}{2} S_{2}(q) \eta^{\prime} S_{2}(q) \right)
\gamma^{\nu} {\cal D}_{\mu\nu}(q-p). \label{bs22-amp}
\ea
Now, we would like to get a solution to these equations. As in the
case of two flavor QCD \cite{bs-short,bs-long}, one could use the
Ward identities to solve the problem.

By comparing the Ward identities in Eqs.~(\ref{P1=0}) and
(\ref{P2=0}) with the pole structure of the vertices in
Eqs.~(\ref{pole0}) and (\ref{pole00}), we derive the required
structure of the BS wave functions in the singlet channel,
\ba
\eta (p) &=& \frac{i}{F^{(\eta)}} \left(
S_{1}^{-1}(p) \gamma^{5} + \gamma^{5} S_{1}^{-1}(p) \right)
=\frac{2}{F^{(\eta)}} \left( \tilde{\Delta}_{1}(p)
{\cal P}_{+} -\Delta_{1}(p) {\cal P}_{-} \right),
\label{eta1-pole} \\
\eta^{\prime} (p) &=& \frac{i}{F^{(\eta)}} \left(
S_{2}^{-1}(p) \gamma^{5} + \gamma^{5} S_{2}^{-1}(p) \right)
=\frac{2}{F^{(\eta)}}
\left( \Delta_{2}(p) {\cal P}_{-}
- \tilde{\Delta}_{2}(p) {\cal P}_{+} \right).
\label{eta2-pole}
\ea
It is straightforward to show that this is indeed the solution to the
set of the BS equations (\ref{bs11-amp}) and (\ref{bs22-amp}),
provided that $\Delta_{1}$ and $\Delta_{2}$ are the solutions to the
gap equations (\ref{D-1}) and (\ref{D-2}).

We use this solution in Sec.~\ref{decay-consts} in order to derive
the decay constants and velocities of the NG bosons.

\subsection{BS equation in the octet channel}
\label{BS-octet}

By following the approach of the previous subsection, let us also
analyze the set of coupled BS equations in the octet channel. Again,
it is clear from the pole structure of the vertices discussed in
Sec.~\ref{Ward-id} that the this channel contains at least one
solution which corresponds to the octet of diquark NG bosons.

We start from establishing the general structure of the matrix BS wave
function. By making use of the definition in Eq.~(\ref{def-pi}), we
obtain the following form of the (non-amputated) wave function in the
channel of interest:
\be
X^{C}_{\pi}(p,P) = \left(\begin{array}{cc}
0 & \delta^{BC} \bfpi_{2}(p,P)  \\
\delta^{AC} \bfpi_{1}(p,P) & d^{ABC} \bfpi_{0}(p,P)
+ if^{ABC} \bfsig (p,P)
\end{array}\right).
\label{mat-BS8wf}
\ee
Notice, that here we added the ``antisymmetric" octet $\bfsig (p,P)$
[see Eq.~(\ref{8times8})] to the general structure because this latter
might in general mix with the NG octet. Of course, in the
weakly coupled limit when the wave function renormalization is close
to 1, the admixture of $\sigma$-octet is expected to be
negligible. The reason is that, according to the (approximate) Ward
identity in Eq.~(\ref{P3=0}), the NG boson does not contain the
antisymmetric contribution proportional to $f^{ABC}$. This is somewhat
similar to the situation with the $\sigma$-singlet in two flavor
QCD \cite{bs-long}.

After substituting the wave function (\ref{mat-BS8wf}) into the BS
equation (\ref{BS-matrix-eq}), we arrive at the following set of
equations:
\ba
S^{-1}_{2}\left(p+\frac{P}{2}\right) \bfpi_{1} (p,P)
S^{-1}_{1}\left(p-\frac{P}{2}\right) &=& \frac{2}{3}
\pi \alpha_{s} \int \frac{d^4 q}{(2\pi )^4} \gamma^{\mu} \left(
\gamma^{5} \bfpi_{2} (q,P) \gamma^{5}
+\frac{5}{\sqrt{6}} \gamma^{5} \bfpi_{0}(q,P) \gamma^{5}
+ 3\sqrt{\frac{3}{2}} \bfsig (q,P) \gamma^{5}
\right) \nonumber\\
&&\times \gamma^{\nu} {\cal D}_{\mu\nu}(q-p), \label{bs1} \\
S^{-1}_{1}\left(p+\frac{P}{2}\right) \bfpi_{2} (p,P)
S^{-1}_{2}\left(p-\frac{P}{2}\right) &=& \frac{2}{3}
\pi \alpha_{s} \int \frac{d^4 q}{(2\pi )^4} \gamma^{\mu} \left(
\gamma^{5} \bfpi_{1} (q,P) \gamma^{5}
+\frac{5}{\sqrt{6}} \gamma^{5} \bfpi_{0}(q,P) \gamma^{5}
- 3\sqrt{\frac{3}{2}} \gamma^{5} \bfsig (q,P)
\right) \nonumber\\
&&\times \gamma^{\nu} {\cal D}_{\mu\nu}(q-p), \label{bs2} \\
S^{-1}_{2}\left(p+\frac{P}{2}\right) \bfpi_{0}(p,P)
S^{-1}_{2}\left(p-\frac{P}{2}\right) &=&
\frac{\pi \alpha_{s}}{2}
\int \frac{d^4 q}{(2\pi )^4} \gamma^{\mu} \Bigg(
2\sqrt{\frac{2}{3}} \gamma^{5} \left[
\bfpi_{1} (q,P) +\bfpi_{2} (q,P) \right] \gamma^{5}
- \gamma^{5} \bfpi_{0}(q,P) \gamma^{5}  \nonumber\\
&&- 3 \bfpi_{0}(q,P) +3 \left[ \bfsig (q,P) \gamma^{5}
- \gamma^{5} \bfsig (q,P) \right] \Bigg)
\gamma^{\nu} {\cal D}_{\mu\nu}(q-p), \label{bs-pi} \\
S^{-1}_{2}\left(p+\frac{P}{2}\right) \bfsig (p,P)
S^{-1}_{2}\left(p-\frac{P}{2}\right) &=&
\frac{\pi \alpha_{s} }{6}
\int \frac{d^4 q}{(2\pi )^4} \gamma^{\mu} \Bigg(
2\sqrt{6} \left[
\bfpi_{1} (q,P) \gamma^{5} -\gamma^{5} \bfpi_{2} (q,P) \right]
+5 \bfpi_{0}(q,P) \gamma^{5} \nonumber\\
&& - 5 \gamma^{5} \bfpi_{0}(q,P)
+5 \gamma^{5} \bfsig (q,P) \gamma^{5} -3 \bfsig (q,P)
\Bigg) \gamma^{\nu} {\cal D}_{\mu\nu}(q-p). \label{bs-sig}
\ea
In deriving these equations, we used the following identities:
\ba
d^{AA^{\prime} B^{\prime}} d^{BB^{\prime} C^{\prime}}
d^{CC^{\prime} A^{\prime}} &=& -\frac{1}{2} d^{ABC} ,\\
d^{AA^{\prime} B^{\prime}} d^{BB^{\prime} C^{\prime}}
f^{CC^{\prime} A^{\prime}} &=& -\frac{5}{6} f^{ABC} ,\\
d^{AA^{\prime} B^{\prime}} f^{BB^{\prime} C^{\prime}}
f^{CC^{\prime} A^{\prime}} &=& -\frac{3}{2} d^{ABC} ,\\
f^{AA^{\prime} B^{\prime}} f^{BB^{\prime} C^{\prime}}
f^{CC^{\prime} A^{\prime}} &=& \frac{3}{2} f^{ABC} .
\ea
Following the same approach as in the case of the singlet channel,
we introduce the amputated wave functions by
\begin{mathletters}
\ba
\pi_{1}(p,P) &=& S^{-1}_{2}\left(p+\frac{P}{2}\right) \bfpi_{1} (p,P)
S^{-1}_{1}\left(p-\frac{P}{2}\right) ,  \label{pi1-amp} \\
\pi_{2}(p,P) &=& S^{-1}_{1}\left(p+\frac{P}{2}\right) \bfpi_{2} (p,P)
S^{-1}_{2}\left(p-\frac{P}{2}\right) , \label{pi2-amp} \\
\pi_{0} (p,P) &=& S^{-1}_{2}\left(p+\frac{P}{2}\right) \bfpi_{0} (p,P)
S^{-1}_{2}\left(p-\frac{P}{2}\right) , \label{pi-amp}\\
\sigma(p,P) &=& S^{-1}_{2}\left(p+\frac{P}{2}\right) \bfsig(p,P)
S^{-1}_{2}\left(p-\frac{P}{2}\right) , \label{sig-amp}
\ea
\label{pi-amp-def}
\end{mathletters}
and consider their BS equations in the limit of a vanishing total
momentum  of the diquark NG bosons, $P \to 0$. Therefore, we arrive
at the following equations:
\ba
\pi_{1} (p)&=& \frac{2}{3}
\pi \alpha_{s} \int \frac{d^4 q}{(2\pi )^4} \gamma^{\mu} \left(
\gamma^{5} S_{1}(q) \pi_{2} (q)
+\frac{5}{\sqrt{6}} \gamma^{5} S_{2}(q) \pi_{0} (q)
+ 3\sqrt{\frac{3}{2}} S_{2}(q) \sigma (q) \right)
S_{2}(q) \gamma^{5} \gamma^{\nu} {\cal D}_{\mu\nu}(q-p),
\label{bs1-amp} \\
\pi_{2} (p)&=& \frac{2}{3}
\pi \alpha_{s} \int \frac{d^4 q}{(2\pi )^4} \gamma^{\mu}
\gamma^{5} S_{2}(q) \left(
\pi_{1} (q) S_{1}(q) \gamma^{5}
+\frac{5}{\sqrt{6}} \pi_{0} (q) S_{2}(q) \gamma^{5}
- 3\sqrt{\frac{3}{2}} \sigma (q) S_{2}(q) \right)
\gamma^{\nu} {\cal D}_{\mu\nu}(q-p), \label{bs2-amp} \\
\pi_{0} (p) &=& \frac{\pi \alpha_{s} }{2}
\int \frac{d^4 q}{(2\pi )^4} \gamma^{\mu} \Bigg(
2\sqrt{\frac{2}{3}} \gamma^{5} \left[
S_{2}(q) \pi_{1} (q) S_{1}(q)
+S_{1}(q) \pi_{2} (q) S_{2}(q) \right] \gamma^{5}
- \gamma^{5} S_{2}(q) \pi_{0} (q) S_{2}(q) \gamma^{5} \nonumber\\
&&- 3 S_{2}(q) \pi_{0} (q) S_{2}(q)
+3 \left[ S_{2}(q) \sigma (q) S_{2}(q) \gamma^{5}
- \gamma^{5} S_{2}(q) \sigma (q) S_{2}(q)
\right] \Bigg) \gamma^{\nu} {\cal D}_{\mu\nu}(q-p),
\label{bs-pi-amp} \\
\sigma (p)&=& \frac{\pi \alpha_{s} }{6}
\int \frac{d^4 q}{(2\pi )^4} \gamma^{\mu} \Bigg(
2\sqrt{6} \left[
S_{2}(q) \pi_{1} (q) S_{1}(q) \gamma^{5}
-\gamma^{5} S_{1}(q) \pi_{2} (q) S_{2}(q) \right]
+5 S_{2}(q) \pi_{0} (q) S_{2}(q) \gamma^{5} \nonumber\\
&&- 5 \gamma^{5} S_{2}(q) \pi_{0} (q) S_{2}(q)
+5 \gamma^{5} S_{2}(q) \sigma (q) S_{2}(q) \gamma^{5}
-3 S_{2}(q) \sigma (q) S_{2}(q)
\Bigg) \gamma^{\nu} {\cal D}_{\mu\nu}(q-p). \label{bs-sig-amp}
\ea
The approximate solution to this set of equations is obtained by
comparing the Ward identities in Eqs.~(\ref{P3=0}), (\ref{P4=0})
and (\ref{P5=0}) with the pole structure of the vertices in
Eqs.~(\ref{pole1}), (\ref{pole2}) and (\ref{pole3}). In this way, we
arrive at the following ansatz:
\ba
\pi_{0} (p) &=& \frac{i}{2F^{(\pi)}} \left(
S_{2}^{-1}(p) \gamma^{5} + \gamma^{5} S_{2}^{-1}(p)\right)
=  \frac{1}{F^{(\pi)}}
\left( \Delta_{2}(p) {\cal P}_{-}
-\tilde{\Delta}_{2}(p) {\cal P}_{+}\right) , \label{pi-pole} \\
\pi_{1} (p) &=& \frac{i}{\sqrt{6}F^{(\pi)}} \left(
S_{2}^{-1}(p) \gamma^{5} + \gamma^{5} S_{1}^{-1}(p)\right)
  = \frac{1}{\sqrt{6}F^{(\pi)}}
\left[ \left( \tilde{\Delta}_{1}(p)
-\tilde{\Delta}_{2}(p)\right) {\cal P}_{+}
+ \left(\Delta_{2}(p)-\Delta_{1}(p) \right) {\cal P}_{-}
\right], \label{pi1-pole} \\
\pi_{2} (p) &=& \frac{i}{\sqrt{6}F^{(\pi)}} \left(
S_{1}^{-1}(p) \gamma^{5} + \gamma^{5} S_{2}^{-1}(p)\right)
  = \frac{1}{\sqrt{6}F^{(\pi)}}
\left[ \left( \tilde{\Delta}_{1}(p)
-\tilde{\Delta}_{2}(p)\right) {\cal P}_{+}
+ \left(\Delta_{2}(p)-\Delta_{1}(p) \right) {\cal P}_{-}
\right], \label{pi2-pole}\\
\sigma (p) &=& 0.\label{sig-pole}
\ea
It is straightforward to show that this is a solution to the set
of the BS equations (\ref{bs1-amp}) -- (\ref{bs-sig-amp}). Notice,
however, that the presented solution is approximate to the same
extent as the quark propagators are. Indeed, if one takes into
account the corrections due to the wave function renormalization
of quarks, the expressions in Eqs.~(\ref{pi1-pole}), (\ref{pi2-pole})
and (\ref{sig-pole}), determined by the Ward identities, would be
modified. It is remarkable that no similar modifications would appear
for the singlet NG boson considered in the previous subsection.
This whole situation resembles quite a lot the analysis in two
flavor dense QCD \cite{bs-long}, where the NG doublets were free
of any admixtures, while the NG singlet had a contribution from
another singlet. All the arguments of Ref.~\cite{bs-long} in support
of the selfconsistency of the leading order approximation
apply without changes to the analysis here.

\section{Decay constants of NG bosons}
\label{decay-consts}

In this section, we derive the values of the decay constants for the NG
bosons in the Pagels-Stokar approximation \cite{PS} (for a review see
Ref.~\cite{Mir}). We start with the definitions in Eqs.~(\ref{decay-1})
and (\ref{decay-8}). It is straightforward, then, to derive the
following
exact expressions:
\ba
P^{(\eta )}_{\mu} F^{(\eta)} &=& \frac{i}{2}
\int \frac{d^{4}q}{(2\pi)^{4}}
\mbox{tr} \left\{\gamma_{\mu} \gamma^{5} \left[
8 \bfeta^{\prime}(q,P) +\bfeta(q,P) \right] \right\} \nonumber\\
&=& \frac{i}{2} \int \frac{d^{4}q}{(2\pi)^{4}}
\mbox{tr} \left\{\gamma_{\mu} \gamma^{5} \left[
8 S_{2}(q+P/2) \eta^{\prime} (q,P) S_{2}(q-P/2)
+ S_{1}(q+P/2) \eta(q,P) S_{1}(q-P/2) \right] \right\},
\label{decay-eta} \\
P^{(\pi)}_{\mu} F^{(\pi)} &=& \frac{i}{12} \int
\frac{d^{4}q}{(2\pi)^{4}} \mbox{tr}
\left[\gamma_{\mu} \gamma^{5} \left( 5 \bfpi_{0}(q,P)
+ \sqrt{6}\bfpi_{1} (q,P) + \sqrt{6}\bfpi_{2} (q,P)
\right) \right] \nonumber\\
&=& \frac{i}{12} \int \frac{d^{4}q}{(2\pi)^{4}} \mbox{tr}
\left[\gamma_{\mu} \gamma^{5}
\left(5 S_{2}(q+P/2) \pi_{0}(q,P) S_{2}(q-P/2)
+ \sqrt{6} S_{2}(q+P/2) \pi_{1} (q,P) S_{1}(q-P/2)
\right. \right. \nonumber \\
&& \left. \left.
+\sqrt{6} S_{1}(q+P/2) \pi_{2} (q,P) S_{2}(q+P/2) \right)
\right].  \label{decay-pi}
\ea
In order to calculate the integrals on right hand side, one needs to know
the explicit form of the BS wave functions at non-zero values of the total
momentum $P$. In general, however, it is hard to derive them. Therefore,
it is natural to consider the Pagels-Stokar approximation \cite{PS,Mir}.
Such an approximation uses the amputated wave functions at zero total
momentum which are given in Eqs.~(\ref{eta1-pole}), (\ref{eta2-pole}) and
in Eqs.~(\ref{pi-pole}), (\ref{pi1-pole}) and (\ref{pi2-pole}) for the
singlet and the octet states, respectively. A simple calculation leads to
the following result [see Eqs.~(\ref{dec-eta}) and (\ref{dec-pi}) in
Appendix A for details]:
\ba
\left(F^{(\eta)}\right)^{2}
\left\{ \begin{array}{c} P_{0} \\ c_{\eta}^{2} \vec{P}
\end{array} \right\}
&\simeq &
\frac{9\mu^{2}}{2\pi^{2}}\left\{ \begin{array}{c} P_{0} \\
\frac{1}{3}\vec{P} \end{array} \right\},
\label{d-eta} \\
\left(F^{(\pi)}\right)^{2}
\left\{ \begin{array}{c} P_{0} \\ c_{\pi}^{2} \vec{P}
\end{array} \right\}
&\simeq &
\frac{\mu^{2}}{24\pi^{2}} \left(7-\frac{|\Delta_{1}^{-}|^2
+|\Delta_{2}^{-}|^2-|\Delta_{1}^{-}-\Delta_{2}^{-}|^2}
{|\Delta_{1}^{-}|^2-|\Delta_{2}^{-}|^2}
\ln\frac{|\Delta_{1}^{-}|^2}{|\Delta_{2}^{-}|^2}\right)
\left\{ \begin{array}{c} P_{0}
\\  \frac{1}{3}\vec{P} \end{array} \right\} \nonumber \\
&\simeq& \frac{(21-8\ln2)\mu^{2}}{72\pi^{2}}
\left\{ \begin{array}{c}
P_{0} \\  \frac{1}{3}\vec{P} \end{array} \right\} ,
\label{d-pi}
\ea
where we used the relation between the gaps $\Delta_{1}^{-}\approx 2
\Delta_{2}^{-}$ as in Eq.~(\ref{D1-D2}).

Recall that here we consider the NG bosons from the left-handed sector.
The expressions for the decay constants of the NG bosons from the
right-handed sector are of course the same. In order to obtain the decay
constants for the scalar and pseudoscalar NG bosons, one has to multiply
the expressions for $(F^{(\eta)})^{2}$ and $(F^{(\pi)})^{2}$ in Eqs.
(\ref{d-eta}) and (\ref{d-pi}) by a factor of $2$.

We see that the decay constants of all NG bosons are of order $\mu$, and
their velocities are equal to $1/\sqrt{3}$. After taking into account the
difference in definitions of the decay constants, we find that expressions
(\ref{d-eta}) and (\ref{d-pi}) agree with those derived in
Ref.~\cite{SonSt}\footnote{Our left-handed and right-handed NG fields
correspond respectively to $X$ and $Y$ fields in the nonlinear realization
of the $SU(3)_c \times SU(3)_L \times SU(3)_R$ symmetry of
Ref.~\cite{SonSt}.} (as well as with those in Ref.~\cite{Zar}, and up to a
factor of 2 in $(F^{(\pi)})^{2}$ with those in Ref.~\cite{Beane}), where a
different approach was used.

A few words should be said about Ref.~\cite{Rho}, also using the method of
the BS equation for studying the NG bosons in the CFL phase of cold dense
QCD. The result for the decay constants obtained there is numerically
different from our result as well as from the results of
Refs.~\cite{SonSt,Zar,Beane}. It seems the reason of that is because,
while being qualitatively correct, the analysis of Ref. \cite{Rho} misses
some relevant quantitative details (for example, no distiction between the
two types of quarks with different gaps seems to be made).

\section{Absence of massive radial excitations}
\label{mass}

The important property of the quark pairing dynamics in cold dense
QCD is the long range interaction mediated by the gluons of the
magnetic type \cite{PR1,Son}. Of course, the Meissner effect would
eventually provide screening for the gluons in the far infrared region.
It is known that such screening is negligible (at least in the leading
order) for the dynamics of the gap formation \cite{us,SW2,PR2,H1,Br1}.
The paring dynamics of the massive diquarks, however, is quite
sensitive to the Meissner effect \cite{bs-short,bs-long}.

Let us now consider the difference between two flavor dense QCD
and three flavor dense QCD. The former has the residual
$SU(2)_{c}$ gauge symmetry in the vacuum, so that three out of total
eight gluons do not feel the Meissner effect. In contrast, the
gauge symmetry of the latter is completely broken (through the Higgs
mechanism). This means, in particular, that all eight gluons in the
CFL phase of three flavor QCD are affected by the Meissner
screening.

The natural question is whether the conjecture of Ref.~\cite{us3}
about the existence of an infinite tower of massive excitations
in the diquark channels with quantum numbers of the NG
bosons could be generalized to three flavor QCD. We recall that
this conjecture was derived from studying some unusual properties
of the effective potential. Taken literally, the conjecture would
imply the existence of an infinite tower of massive excitations for
each of the five (would be) NG bosons in the case of QCD with two
flavors. However, our detailed analysis \cite{bs-short,bs-long} shows
that, in fact, an infinite tower of the excitations occurs only in
the color singlet channel. No radial excitations of the four
other (would be) NG bosons (from a color doublet and antidoublet)
appear. In order to reach this conclusion, one needs to take into
account the effect of the Meissner screening for the pairing dynamics
of diquark bound states.

By following the same arguments as in the case of two flavor QCD
\cite{bs-short,bs-long}, one should distinguish between two classes of
bound states, for which the role of the Meissner effect is very different.
The first class consists of light bound states with the masses $M \ll
|\Delta_{0}^{-}|$. The binding energy of these states is large (tightly
bound states), and the Meissner effect is essentially irrelevant for their
pairing dynamics. This point could be illustrated by the BS equations for
the lightest diquarks, the massless NG bosons: in that case the most
important region of momenta in the equations is given by $|\Delta^{-}_{0}|
\alt |k_{0}| \ll |\vec{k}| \alt \mu$ where the Meissner effect is
negligible (see Sec.~VII and Appendix~C in Ref.~\cite{bs-long}).

The second class includes quasiclassical states with the masses close to
their threshold. Since the binding energy of the quasiclassical states is
small, the quasiclassical part of the spectrum is almost completely
determined by the behavior of the potential at large distances.  In the
particular case of cold dense QCD, the interaction between quarks is long
ranged in the (imaginary) time direction and essentially short ranged in
the spatial ones \cite{Son,us3}.  Because of that, the far infrared
region, with $|k_{0}| < |\Delta^{-}_{0}| \alt |\vec{k}|$, is particularly
important for the pairing dynamics of the quasiclassical diquark states
and the Meissner effect is rather strong in that region. This implies that
the inclusion of the Meissner effect is crucial for extracting the
properties of the states from this second class (for more details see
Sec.~VII and Appendix~C in Ref.~\cite{bs-long}).

Now, by repeating the arguments of Ref.~\cite{bs-long}, we conclude that,
because of the Meissner effect, an infinite tower of (quasiclassical)
massive diquarks does not appear in the CFL phase of three flavor QCD.
This is directly related to the fact that all eight gluons in the model at
hand are affected by the Meissner screening. This qualitative argument by
itself does not prevent the possibility of a finite number of radial
excitations in the spectrum. However, the same type of analysis as in
Ref.~\cite{bs-long} shows that no radial excitations of NG bosons appear
at all. After the Meissner effect is taken into account, the interaction
provided by gluons appears to be too weak to form even the lowest massive
radial excitation of a NG boson.

\section{Conclusions}
\label{conclusion}

In this paper we have studied the properties of diquark states with the
quantum numbers of the NG bosons in the CFL phase of cold dense QCD with
three quark flavors. We have derived the general Bethe-Salpeter equations
in the singlet and the octet channels that include all the NG bosons.

Our analytical analysis of the Bethe-Salpeter equations in the CFL phase
shows that the theory contains one pseudoscalar octet of NG bosons, one
singlet NG boson and one singlet pseudo-NG boson. This agrees with the
previous results obtained in the effective theory approach of
Refs.~\cite{CasGat,SonSt,Rho,HZB,Zar,Beane}. We calculate the decay
constants of the (pseudo-) NG bosons by using the Pagels-Stokar
approximation, and the results agree with those of Ref.~\cite{SonSt},
obtained by a different method.

We also show that there are no spin zero massive states (with quantum
numbers of the NG bosons) in the CFL phase of cold dense QCD. This
conclusion is reached by studying a specific realization of the Meissner
effect in three flavor QCD. In contrast to the case of dense QCD with two
flavors, all eight gluons in the CFL phase become massive due to the
Meissner effect. As a result, the one-gluon interaction becomes rather
weak in the far infrared region which is responsible for the pairing
dynamics of the (would be quasiclassical) massive radial excitations of
the NG bosons \cite{bs-long}. We note, though, that our current
conclusions state nothing about the bound diquark states of higher spins.
The problem of the higher spin channels might be also interesting, but it
is beyond the scope of this paper.

\begin{acknowledgments}
I.A.S. would like to thank M.~Stephanov and D.T.~Son for valuable
comments.  V.A.M. is grateful to Hiroshi Toki for his warm
hospitality at Osaka University. L.C.R.W. thanks Professor
K.~Tennakone for his  hospitality and support at the IFS.
The work of V.A.M. was partly supported by the
Grant-in-Aid of Japan Society for the Promotion of Science No.~11695030.
The work of I.A.S. was supported by the U.S. Department of Energy Grants
No.~DE-FG02-84ER40153 and No.~DE-FG02-87ER40328. The work of L.C.R.W. was
supported by the U.S. Department of Energy Grant No.~DE-FG02-84ER40153.

\end{acknowledgments}

\appendix

\section{Calculation of the NG boson decay constants}

In this appendix we exhibit in detail our calculation of the NG decay
constants. We start from the definitions in Eqs.~(\ref{decay-eta}) and
(\ref{decay-pi}). In the Pagels-Stokar approximation, the BS wave
functions of the NG bosons are taken at vanishing total momentum $P$.
The explicite form of the wave functions in the singlet channel is
given in Eqs.~(\ref{eta1-pole}) and (\ref{eta2-pole}). By making use of
these expressions, we rewrite the right hand side of
Eq.~(\ref{decay-eta})
as follows:
\ba
P^{(\eta )}_{\mu} F^{(\eta)} &\simeq& \frac{i}{2F^{(\eta)}}
\int \frac{d^{4}q}{(2\pi)^{4}}
\mbox{tr} \left\{ 8 \gamma_{\mu} S_{2}(q) \Pslash \left[S_{2}(q)
+\gamma^{5} S_{2}(q) \gamma^{5} \right] + \gamma_{\mu} S_{1}(q)
\Pslash \left[S_{1}(q) +\gamma^{5} S_{1}(q) \gamma^{5} \right]
\right\} +O \left(P^{2}\right)\nonumber\\
&\simeq& \frac{2}{F^{(\eta)}} P^{\nu}
\int\frac{d q_{4} d^{3} \vec{q}}{(2\pi)^{4}}
\left(g_{\mu 0} g_{\nu 0}
-\frac{\vec{q}_{\mu} \vec{q}_{\nu}}{|\vec{q}|^{2}}\right)
\left( \frac{8|\Delta^{-}_{2}|^{2}}{\left[ q_{4}^{2}
+(\epsilon_{q}^{-})^{2} + |\Delta^{-}_{2}|^{2} \right]^2}
+\frac{|\Delta^{-}_{1}|^{2}}{\left[ q_{4}^{2}
+(\epsilon_{q}^{-})^{2} + |\Delta^{-}_{1}|^{2} \right]^2}
\right) +O \left(P^{2}\right)\nonumber\\
&\simeq& \frac{9\mu^{2}}{2\pi^{2} F^{(\eta)}}\left(g_{\mu 0} P_{0}
+\frac{1}{3} \vec{P}_{\mu} \right) +O \left(P^{2}\right),
\label{dec-eta}
\ea
where we used the expansion of the quark propagators in
powers of momentum $P$,
\be
S_{1,2}(q+P/2) \simeq S_{1,2}(q)
+ \frac{i}{2} S_{1,2}(q) \Pslash S_{1,2}(q)
+O \left(P^{2}\right),
\ee
as well as the following results for the Dirac traces:
\ba
\mbox{tr} \left( \gamma_{\mu} \Lambda_{q}^{\pm} \gamma_{\nu}
\Lambda_{q}^{\mp} {\cal P}_{\mp} \right)
= g_{\mu 0} g_{\nu 0}
-\frac{\vec{q}_{\mu} \vec{q}_{\nu}}{|\vec{q}|^{2}}.
\ea
In the case of the octet state, the calculation is similar. The
explicite form of the corresponding BS wave functions is given
in Eqs.~(\ref{pi-pole}), (\ref{pi1-pole}) and (\ref{pi2-pole}).
By substituting them into Eq.~(\ref{decay-pi}), we arrive at
\ba
P^{(\pi)}_{\mu} F^{(\pi)} &\simeq& \frac{i}{24 F^{(\pi)}}
\int \frac{d^{4}q}{(2\pi)^{4}}
\mbox{tr} \Bigg\{ 5 \gamma_{\mu} S_{2}(q) \Pslash \left[S_{2}(q)
+\gamma^{5} S_{2}(q) \gamma^{5} \right] \nonumber \\
&+& 2 \gamma_{\mu} \left[ S_{1}(q)
+ \gamma^{5} S_{2}(q) \gamma^{5} \right] \Pslash S_{1}(q)
+ 2 \gamma_{\mu} \left[ S_{2}(q)
+ \gamma^{5} S_{1}(q) \gamma^{5} \right] \Pslash S_{2}(q)
\Bigg\} +O \left(P^{2}\right)\nonumber\\
&\simeq&  \frac{1}{6 F^{(\pi)}}  P^{\nu}
\int\frac{d q_{4} d^{3} \vec{q}}{(2\pi)^{4}}
\left(g_{\mu 0} g_{\nu 0}
-\frac{\vec{q}_{\mu} \vec{q}_{\nu}}{|\vec{q}|^{2}}\right)
\Bigg( \frac{5|\Delta^{-}_{2}|^{2}}{\left[ q_{4}^{2}
+(\epsilon_{q}^{-})^{2} + |\Delta^{-}_{2}|^{2} \right]^2}
+\frac{|\Delta^{-}_{1}|^{2}}{\left[ q_{4}^{2}
+(\epsilon_{q}^{-})^{2} + |\Delta^{-}_{1}|^{2} \right]^2}
\nonumber\\
&+& \frac{|\Delta^{-}_{2}|^{2}}{\left[ q_{4}^{2}
+(\epsilon_{q}^{-})^{2} + |\Delta^{-}_{2}|^{2} \right]^2}
-\frac{\Delta^{-}_{1}(\Delta^{-}_{2})^{*}
+\Delta^{-}_{2}(\Delta^{-}_{1})^{*} }{
\left[ q_{4}^{2}
+(\epsilon_{q}^{-})^{2} + |\Delta^{-}_{1}|^{2} \right]
\left[ q_{4}^{2}
+(\epsilon_{q}^{-})^{2} + |\Delta^{-}_{2}|^{2} \right]}
\Bigg) +O \left(P^{2}\right)\nonumber\\
&\simeq& \frac{\mu^{2}}{24\pi^{2} F^{(\pi)}}
\left(7-\frac{\Delta^{-}_{1}(\Delta^{-}_{2})^{*}
+\Delta^{-}_{2}(\Delta^{-}_{1})^{*} }
{|\Delta^{-}_{1}|^{2}-|\Delta^{-}_{2}|^{2}}
\ln\frac{|\Delta^{-}_{1}|^{2}}{|\Delta^{-}_{2}|^{2}}
\right) \left(g_{\mu 0}
P_{0} +\frac{1}{3} \vec{P}_{\mu} \right) +O \left(P^{2}\right).
\label{dec-pi}
\ea


\begin{references}

\bibitem{W1} M.~Alford, K.~Rajagopal, and F.~Wilczek,
\pl B{\bf 422}, 247 (1998).

\bibitem{S1} R.~Rapp, T.~Sch\"{a}fer, E.V.~Shuryak, and
M.~Velkovsky, \prl {\bf 81}, 53 (1998).

\bibitem{BarFra} B.C.~Barrois, Nucl. Phys. {\bf B129}, 390 (1977);
S.C.~Frautschi, in ``Hadronic matter at extreme energy density",
edited by N.~Cabibbo and L.~Sertorio (Plenum Press, 1980).

\bibitem{Bail} D.~Bailin and A.~Love,
Nucl. Phys. {\bf B190}, 175 (1981);
Nucl. Phys. {\bf B205}, 119 (1982);
Phys. Rep. {\bf 107}, 325 (1984).

\bibitem{PR1} R.D.~Pisarski and D.H.~Rischke,
\prl {\bf 83}, 37 (1999).

\bibitem{Son} D.T.~Son, \prd {\bf 59}, 094019 (1999).

\bibitem{us} D.K.~Hong, V.A.~Miransky, I.A.~Shovkovy,
and L.C.R.~Wijewardhana, \prd {\bf 61}, 056001 (2000).

\bibitem{SW2} T.~Sch\"{a}fer and F.~Wilczek,
\prd {\bf 60}, 114033 (1999).

\bibitem{PR2} R.D.~Pisarski and D.H.~Rischke,
\prd {\bf 61}, 051501 (2000).

\bibitem{H1} S.D.H.~Hsu and M.~Schwetz,
Nucl. Phys. {\bf B572}, 211 (2000).

\bibitem{Br1} W.E.~Brown, J.T.~Liu, and H.-C.~Ren,
\prd {\bf 61}, 114012 (2000); {\em ibid.} {\bf 62}, 054016 (2000).

\bibitem{CFL} M.~Alford, K.~Rajagopal, and F.~Wilczek,
Nucl. Phys. {\bf B537}, 443 (1999); M.~Alford, J.~Berges,
and K.~Rajagopal, Nucl. Phys. {\bf B558}, 219 (1999).

\bibitem{cnt} T.~Sch\"{a}fer and F.~Wilczek, \prl
{\bf 82}, 3956 (1999); \prd {\bf 60}, 074014 (1999).

\bibitem{us3} V.A.~Miransky, I.A.~Shovkovy, and
L.C.R.~Wijewardhana, \pl B {\bf 468}, 270 (1999).

\bibitem{us2} I.A.~Shovkovy and L.C.R.~Wijewardhana,
\pl B {\bf 470}, 189 (1999).

\bibitem{Spt} T.~Sch\"{a}fer,
Nucl. Phys. {\bf B575}, 269 (2000).

\bibitem{013} S.R.~Beane, P.F.~Bedaque, and M.J.~Savage,
nucl-th/0004013.

\bibitem{Br2} W.E.~Brown, J.T.~Liu, and H.-C.~Ren,
\prd {\bf 62}, 054013 (2000).

\bibitem{Rsch} D.H.~Rischke, \prd {\bf 62}, 034007 (2000);
{\em ibid.} {\bf 62}, 054017 (2000).

\bibitem{CaD} G.W.~Carter and D.~Diakonov,
Nucl. Phys. {\bf B582}, 571 (2000);
R.~Casalbuoni, Z.~Duan, and F.~Sannino,
\prd {\bf 62}, 094004 (2000).

\bibitem{bs-short} V.A.~Miransky, I.A.~Shovkovy, and
L.C.R.~Wijewardhana, hep-ph/0003327.

\bibitem{bs-long} V.A.~Miransky, I.A.~Shovkovy, and
L.C.R.~Wijewardhana, \prd {\bf 62}, 085025 (2000).

\bibitem{CasGat} R.~Casalbuoni and R.~Gatto,
\pl B {\bf 464}, 111 (1999); hep-ph/9911223;
D.K.~Hong, M.~Rho, and I.~Zahed, \pl B {\bf 468}, 261 (1999).

\bibitem{SonSt} D.T.~Son and M.A.~Stephanov,
\prd {\bf 61}, 074012 (2000);
{\em ibid.} {\bf 62}, 059902(E) (2000).

\bibitem{Rho}  M.~Rho, A.~Wirzba, and I.~Zahed,
\pl B {\bf 473}, 126 (2000);
M.~Rho, E.~Shuryak, A.~Wirzba, and I.~Zahed,
Nucl. Phys. {\bf A676}, 273 (2000).

\bibitem{HZB} D.K.~Hong, T.~Lee, and D.-P.~Min,
\pl B {\bf 477}, 137 (2000);
C.~Manuel and M.G.H.~Tytgat,
{\em ibid.} {\bf 479}, 190 (2000).

\bibitem{Zar} K.~Zarembo, \prd {\bf 62}, 054003 (2000).

\bibitem{Beane} S.R.~Beane, P.F.~Bedaque, and M.J.~Savage,
\pl B {\bf 483}, 131 (2000).

\bibitem{JJ} R. Jackiw and K. Johnson, \prd {\bf 8}, 2386
(1973);
J.M. Cornwall and R.E. Norton, {\em ibid.} {\bf 8}, 3338
(1973).

\bibitem{PS} H.~Pagels and S.~Stokar, \prd {\bf 20}, 2947 (1979).

\bibitem{Mir} V.A.~Miransky, {\em Dynamical Symmetry Breaking in
Quantum Field Theories} (World Scientific, Singapore, 1993).

\end{references}
\end{document}